\documentclass[12pt]{article}%
\usepackage{amsmath}
\usepackage{amssymb}
\usepackage{amsfonts}
\usepackage{graphicx}%
\input epsf.tex
\newcommand{\be}{\begin{equation}}
\newcommand{\ee}{\end{equation}}
\newcommand{\bea}{\begin{eqnarray}}
\newcommand{\eea}{\end{eqnarray}}

\begin{document}
\bigskip\begin{titlepage}
\begin{flushright}
UUITP-10/12\\
\end{flushright}
\vspace{1cm}
\begin{center}
{\Large\bf Constraining dark energy\\}
\end{center}
\vspace{3mm}
\begin{center}
{\large
Ulf H.\ Danielsson} \\
\vspace{5mm}
Institutionen f\"or fysik och astronomi, Uppsala Universitet, \\
Box 516, SE-751 20
Uppsala, Sweden\\
\vspace{5mm}
{\tt
ulf.danielsson@physics.uu.se \\
}
\end{center}
\vspace{5mm}
\begin{center}
{\large \bf Abstract}
\end{center}
In this paper we propose a mechanism that protects theories violating a holographic bound suggested in arXiv:1203.5476
from developing accelerated expansion. The mechanism builts on work on transplanckian physics, and a non-trivial choice of vacuum
states. If correct, it lends further support for detectable signatures in the CMBR signalling new physics.
\vfill
\begin{flushleft}
\end{flushleft}
\end{titlepage}\newpage


\section{Introduction}

\bigskip

It is an intriguing and challenging problem to fit a cosmology undergoing
accelerated expansion into a theory of strings or quantum gravity. It is also
a problem of the utmost physical importance, since it is now an established
fact that the universe not only underwent accelerated expansion during its
early days, i.e. inflation, but is again entering into such a phase during the
last few billion years. Apart from the challenges of detailed de Sitter model
building in string theory, see \cite{Burgess} for a recent review and
references, there are deep problems related to the concept of entropy in
gravity and the possible application of holography. While black holes and anti
de Sitter spaces are reasonably well understood from this point of view, de
Sitter space remain elusive.

There has been numerous, more or less successful, attempts to apply holography
to cosmology. In \cite{Danielsson 2003} an attempt was made to clear up some
of the associated issues. In particular, it was argued that the de Sitter
horizon, in analogue with a black hole horizon, should be thought as
describing entropy of matter \textit{beyond} the horizon. Or, to be more
precise, matter that has passed out through the horizon. Hence, the horizon
entropy can only indirectly be used to limit the property of matter inside of
the de Sitter radius. This is achieved through the study of how the area of
the horizon changes in response to the flow of matter.

Recently, it was suggested in \cite{Conlon: 2012} that holography actually do
provide interesting constraints on models of dark energy and accelerated
expansion. The idea is to count the number of initial field configurations
that eventually develop into asymptotic de Sitter space, and compare this
number with the holographic bound in de Sitter space. The relevant entropy
obtained in this way is argued to be of the order%
\begin{equation}
S\sim\frac{Na_{\min}^{-2}}{H^{2}},\label{asymptds}%
\end{equation}
where $a_{\min}$ is a cutoff associated with the light fields of the theory,
and $N$ is the number of such fields. According to holography this number
should be less than%
\begin{equation}
S_{dS}=\frac{8\pi^{2}M_{pl}^{2}}{H^{2}}.
\end{equation}
There is an uncertainty when it comes to the actual numerical coefficient in
the calculation leading to (\ref{asymptds}). One specific possibility is,
according to \cite{Conlon: 2012}, that the number coincides with the
entanglement entropy. Using this, and for the sake of
definiteness, the regularization proposed by \cite{Srednicki}, we find that%
\begin{equation}
0.3\frac{Na_{\min}^{-2}}{H^{2}}<\frac{8\pi^{2}M_{pl}^{2}}{H^{2}},
\end{equation}
which leads to%
\begin{equation}
N\frac{a_{\min}^{-2}}{M_{pl}^{2}}\lesssim27\pi^{2}.
\end{equation}
In \cite{Conlon: 2012} it was specifically argued that there exists in any
typical string theory realization of accelerated expansion, a large number of
axionic fields with periodicity
\begin{equation}
\phi_{a}\rightarrow\phi_{a}+f_{a}.
\end{equation}
It was furthermore argued that the relevant cutoff in the entropy calculation
is provided by the periodicity. To be precise,
\begin{equation}
a_{\min}=\frac{1}{f_{a}\sqrt{2}}.
\end{equation}
According to \cite{Conlon: 2012} these results provide highly nontrivial
constraints on stringy realizations of inflation and dark energy.

While these results are quite intriguing, there are several questions that
remain to be answered. I particular, it would be interesting to see how this
limit is actually implemented by quantum gravity. What would happen if we pick
a theory with the potential to violate the holographic bound, and force the
theory into a state with a positive dark energy? How does the inconsistency
show up? Perhaps there is some kind of physical mechanism that prevents the
system from ending up in the holographically forbidden state? It turns out
that the answer to this question can be found in a surprising place. Let us
turn to the problem of transplanckian physics and its effects on the CMBR.

\section{Transplanckian physics and the CMBR}

\bigskip

As argued in, e.g., \cite{BM} and further discussed from various aspects in
works such as [6-17] (this is just a selected few), it can be expected that
physics beyond the string or Planck scale is magnified through the expansion
of the universe, and affects phenomena at lower energies such as the spectrum
of the CMBR. In inflationary model building it is usually assumed that the
vacuum to be used for all the relevant fields is the Bunch-Davies vacuum. This
is a reasonable choice if all modes can be traced back to infinitely small
scales, where the expansion of the universe, and the resulting deviation from
Minkowsky space, become negligible. However, in the presence of a fundamental
energy scale such as the string or Planck scale there is no reason to expect
the Bunch-Davies vacuum to be the preferred choice, \cite{Danielsson:2002kx}.
New physics, connected with the string or Planck scale, can be expected to
modify the vacuum. This can be conveniently modelled through a Bogolubov
mixing linear in $\frac{H}{\Lambda},$ where there is a dependence on scale
only through the Hubble constant $H$. $\Lambda$ is the energy scale of the new
physics, which could be the string scale or the Planck scale.

As argued in \cite{Danielsson:2005cc}, the effects propagating down to low
energy will be of two types: a modulation of the CMBR spectrum and a back
reaction on the expansion of the universe. The latter effect is the one that
will be of relevance to us.

\bigskip

\subsection{Effects on the CMBR}

\bigskip

According to the analysis of \ \cite{Danielsson:2002kx}, given a Bogolubov
mixing as described above, the typical effect to be expected on the primordial
spectrum is of the form%

\begin{equation}
P(k)=\left(  \frac{H}{\overset{\cdot}{\phi}}\right)  ^{2}\left(  \frac{H}%
{2\pi}\right)  ^{2}\left(  1-\frac{H}{\Lambda}\sin\left(  \frac{2\Lambda}%
{H}\right)  \right)  ,
\end{equation}
where we note a characteristic, relative amplitude of the correction given by
$\frac{H}{\Lambda}$, and a modulation sensitively depending on how
$\frac{\Lambda}{H}$ changes with $k$. The claim is that whatever the nature of
the high energy physics really is, a modulated spectrum of this form is what
we should naturally expect. In \cite{Bergstrom:2002yd} one can find an early
discussion of the phenomenological relevance of the effect, and how its
magnitude is related to the characteristic parameters describing the
inflationary phase. Using the standard slow roll approximation, where an
important parameter is
\begin{equation}
\varepsilon=\frac{M_{pl}^{2}}{2}\left(  \frac{V^{\prime}}{V}\right)
^{2},\label{epsV}%
\end{equation}
with initial conditions imposed at some fundamental scale $\Lambda=\gamma
M_{pl}$, it is found that
\begin{equation}
\frac{\Delta k}{k}\sim\frac{\pi H}{\varepsilon\Lambda}\sim1.3\cdot10^{-3}%
\frac{1}{\gamma\sqrt{\varepsilon}},
\end{equation}
and
\begin{equation}
\frac{H}{\Lambda}\sim4\cdot10^{-4}\frac{\sqrt{\varepsilon}}{\gamma
}.\label{eq:effect}%
\end{equation}
These two relations are the key to estimating the expected magnitude of the
effect. For instance, with a string scale a couple of order of magnitudes
below the Planck scale, and a slow roll parameter $\varepsilon\sim10^{-2}$, we
find an amplitude of $\frac{H}{\Lambda}\sim10^{-2}$ -- comparable with cosmic
variance -- and a periodicity given by $\frac{\Delta k}{k}\sim\mathcal{O}%
\left(  1\right)  $.

\bigskip

\subsection{Back reaction}

\bigskip

The presence of a non-standard vacuum, motivated by the presence of unknown
high energy physics, raises the issue of backreaction. Focusing on the
contribution to the vacuum energy coming form the non-standard vacuum, as
compared with the Bunch-Davies vacuum, one finds an additional energy density
naively given by $\rho_{\Lambda}\sim\Lambda^{2}H^{2}$. To lowest order, as
long as $\Lambda\ll M_{p}$, we can ignore this contribution as was concluded
in \cite{Tanaka:2000jw}. In \cite{Danielsson:2004xw} and
\cite{Danielsson:2005cc}, however, the discussion was taken a step further and
it was noted that the presence of the background energy will change the
effective slow roll parameters. In fact, there are regimes where these effects
even will dominate.

To proceed with a quantitative analysis, we denote the slow roll parameter
describing the time dependence of the Hubble constant by $\varepsilon$ and
define it through%
\begin{equation}
\varepsilon=\frac{\dot{H}}{H^{2}}.\label{epshprick}%
\end{equation}
In addition, we introduce a second slow roll parameter governing the rolling
of the inflaton according to%
\begin{equation}
\varepsilon_{\inf}=\frac{\dot{\phi}^{2}}{2M_{pl}^{2}H^{2}},\label{einf}%
\end{equation}
where $\phi$ is a canonically normalized inflaton. In the slow role
approximation, and in the absence of back reaction from the vacuum, we would
have had $\varepsilon=\varepsilon_{\inf}$, also coinciding with (\ref{epsV}).
With back reaction, as explained in \cite{Danielsson:2005cc}, we find a
decoupling of the expressions for the amplitude and the period according to
\begin{equation}
\frac{H}{\Lambda}\sim4\cdot10^{-4}\frac{\sqrt{\varepsilon_{\inf}}}{\gamma},
\end{equation}
and%
\begin{equation}
\frac{\Delta k}{k}\sim\frac{\pi H}{\varepsilon\Lambda}\sim1.3\cdot10^{-3}%
\frac{\sqrt{\varepsilon_{\inf}}}{\gamma\varepsilon}.\label{eq:period}%
\end{equation}
Let us now proceed with an estimate of $\varepsilon$, following
\cite{Danielsson:2004xw}.

The above estimate of the energy density is not good enough when we want to
find an expression for $\varepsilon$. What we need to do is to take into
account that $H$ will be changing with time, i.e. decrease. Modes with low
momenta were created at earlier times when the value of $H$ were larger, and
the contribution to the energy density from these modes will, as a
consequence, be enhanced. We therefore find an energy density given by
\begin{equation}
\rho_{\Lambda}\left(  a\right)  =\frac{1}{2\pi^{2}}\int_{\varepsilon}%
^{\Lambda}dpp^{3}\frac{H^{2}\left(  \frac{ap}{\Lambda}\right)  }{\Lambda^{2}%
}=\frac{1}{2\pi^{2}}\frac{\Lambda^{2}}{a^{4}}\int_{a_{i}}^{a}daa^{3}%
H^{2}\left(  a\right)  ,
\end{equation}
where we have introduced a low energy cutoff corresponding to the energy at
the time of observation of modes that started out at $\Lambda$ at some
arbitrary initial scale factor $a_{i}$. If we take a derivative of the energy
density with respect to the scale factor and use $\frac{d}{da}=\frac{1}%
{aH}\frac{d}{dt}$, we find%
\begin{equation}
\dot{\rho}_{\Lambda}+4H\rho_{\Lambda}=\frac{1}{2\pi^{2}}\Lambda^{2}H^{3},
\label{eq:contQ}%
\end{equation}
from which we conclude that we must introduce a source term in the Friedmann
equations. It was found in \cite{Danielsson:2004xw} that the evolution is
governed by%
\begin{equation}
\frac{d}{da}\left(  a^{5}HH^{\prime}\right)  =-\frac{\Lambda^{2}}{3\pi
^{2}M_{pl}^{2}}a^{3}H^{2}-\frac{1}{2M_{pl}^{2}}\frac{d}{da}\left(
a^{4}\left(  aH\phi^{\prime}\right)  ^{2}\right)  ,
\end{equation}
where we let $%
\acute{}%
=\frac{d}{da}$. The first term on the right hand side is due to the presence
of the non-standard vacuum, while the second term is due to the presence of
the inflaton potential. In a situation where the first term dominates, we find
a slow roll governed by
\begin{equation}
\varepsilon=\frac{\gamma^{2}}{12\pi^{2}},
\end{equation}
for small $\gamma=\frac{\Lambda}{M_{pl}}$.\footnote{In this paper we
consistently use the reduced Planck mass $M_{pl}=1/\sqrt{8\pi G}\sim
2.4\cdot10^{18}\mathrm{GeV}$. This should be kept in mind when comparing with
the results in \cite{Danielsson:2005cc}.} In the standard case, with no vacuum
contribution, the only non-vanishing term is the second one, leading to a slow
roll governed by (\ref{einf}).

While it is the slow roll of the inflaton that controls the overall amplitude
of the primordial spectrum, one could easily imagine, as discussed in
\cite{Danielsson:2005cc}, that there are more fields in the non-standard
vacuum. These would also contribute to the back reaction and enhance
$\varepsilon$ by a factor $N$, where $N$ is the number of participating
fields. Hence, the expression becomes%
\begin{equation}
\varepsilon=\frac{N\gamma^{2}}{12\pi^{2}}.\label{eq:ebak}%
\end{equation}
We will not discuss the further application of these results to the CMBR, but
instead focus on their relevance to the problem of holographic constraints on
dark energy.

\bigskip

\section{The decay of dark energy}

\bigskip

Let us now make the connection with the holographic constraint on dark energy.
In \cite{Conlon: 2012} it was argued that theories with too many fields for a
given cutoff break the holographic bound, and are inconsistent as models of de
Sitter space. Using the results of the previous section we can now propose a
mechanism for how this comes about.

The key is the decay of the dark energy that results if the vacuum of the
theory is different from the Bunch-Davies vacuum. The calculated value of the
slow roll parameter $\varepsilon$ depends on the number of fields and the
cutoff. If we choose  the number of fields in the new vacuum, and the cutoff
at $\Lambda=1/a_{\min}$, such that the holographic bound is violated, we find
that the slow roll parameter becomes%
\[
\varepsilon>\frac{27\pi^{2}}{12\pi^{2}}\gtrsim1.
\]
This implies that the dark energy decays so fast that no accelerated expansion
will take place, and, therefore, there cannot arise any conflict with the
holographic de Sitter bound. It should be noted that theories that threatens
to violate the bound are not necessarily inconsistent. It is just that these
theories lack, when quantum gravity is taking into account, solutions
corresponding to accelerated expansion.

The proposed mechanism relies on one assumption: \textit{in the presence of
\ a fundamental scale due to stringy or quantum gravitational effects, there
is no reason, in an expanding cosmology, to prefer the Bunch-Davies vacuum. A
much more natural candidate is a vacuum such as the instantaneous Minkowsky
vacuum, as proposed in \cite{Danielsson:2002kx}.} Given this, the theory
automatically protects itself from de Sitter vacua in cases where the de
Sitter bound may be violated. In theories where the bound is not violated,
accelerated expansion is allowed but the decay of the dark energy due to the
vacuum effect may still be physically relevant, and could play an important
role during, e.g., inflation.

The argument can also be turned around. The inconsistency of accelerated
expansion in certain models suggest the need for a general mechanism of vacuum
decay. Such a mechanism is supplied if the natural vacuum choice of quantum
gravity is not the Bunch-Davies. In fact, it turns out that the vacuum choice
of \cite{Danielsson:2002kx} has precisely the desired effect. This, it can be
argued, gives independent support for the idea of detectable, transplanckian
signatures in the CMBR.

\bigskip

\section*{Acknowledgments}

The work was supported by the Swedish Research Council (VR), and the G\"{o}ran
Gustafsson Foundation.

\end{document}